\documentclass[preprint,12pt,a4paper]{elsarticle}
\usepackage{graphicx,amsmath,amssymb,booktabs,array,tikz}
\usetikzlibrary{calc,arrows.meta}
\usepackage[margin=25mm]{geometry}
\usepackage[hidelinks]{hyperref}
\biboptions{sort&compress}
\journal{Nuclear Instruments and Methods in Physics Research Section A}

\makeatletter
\def\ps@pprintTitle{%
  \let\@oddhead\@empty
  \let\@evenhead\@empty
  \def\@oddfoot{\hfil\thepage\hfil}%
  \let\@evenfoot\@oddfoot}
\makeatother

\begin{document}

\begin{frontmatter}

\title{Equivalent-circuit analysis of the transverse resonant beam coupling impedance due to ceramic-chamber RF shields in the J-PARC Main Ring spill-feedback quadrupoles}

\author{Masahito Tomizawa\corref{cor1}}
\ead{masahito.tomizawa@kek.jp}
\cortext[cor1]{Corresponding author}
\affiliation{organization={High Energy Accelerator Research Organization (KEK) / J-PARC Center},
  city={Tsukuba and Tokai}, country={Japan}}

\begin{abstract}
Ceramic vacuum chambers in fast-ramping or fast-response magnets are equipped with RF shields of conducting strips closed through lumped
capacitors, which form a resonant circuit coupled to the transverse dipole moment of the beam. We evaluate the transverse beam coupling impedance
due to the RF shields of the ceramic chambers in the spill-feedback quadrupole magnets of the J-PARC Main Ring. The RF shield is modelled by a
magneto-quasi-static partial element equivalent circuit, and the laminated yoke of the magnet by a two-dimensional field solution with a complex permeability.
The analysis shows a low-$Q$ resonant coupling impedance. The yoke lowers the resonance frequency and raises the shunt impedance per quality factor, and its effect
cannot be neglected. The quality factor is determined mainly by the losses in the capacitors and in the laminations.
The predicted peak impedance of the three chambers exceeds the resistive-wall impedance of the whole ring. The model
can be applied to other machines and to the design of future rapid-cycling synchrotrons.
\end{abstract}

\begin{keyword}
RF shield \sep ceramic chamber \sep transverse impedance \sep PEEC \sep laminated yoke \sep slow extraction
\end{keyword}

\end{frontmatter}

\section{Introduction}

Ceramic (alumina) vacuum chambers are used in magnets whose field changes rapidly, such as the main magnets of rapid cycling
synchrotrons and fast-response correctors, since a metallic chamber would carry large eddy currents. To provide a low-impedance
path for the beam image current, the ceramic is covered with an RF shield (screen) consisting of longitudinal conducting strips or wires.
To interrupt the eddy-current loop for the time-varying magnetic field, each strip is connected to the end flange or ring through a capacitor
at one end. The capacitors are chosen to have a low impedance for the beam and a high impedance at the frequencies of the magnetic field.

The strips, the capacitors and the rings form a resonant circuit. For the transverse (dipole) mode, in which the strips carry
currents ideally proportional to $\cos\theta$, the circuit is a single-turn loop closed through the capacitors, and it couples to the beam
as the secondary of a transformer. At ISIS, probe-coil measurements on a spare quadrupole magnet showed a broad resonant transverse impedance $Z_\perp$ at
185~kHz, and such resonances of the RF screens in the dipole and quadrupole magnets were identified as the main driver of the vertical
head-tail instability~\cite{HB23,IPAC26ISIS}. At the rapid cycling synchrotron of the China Spallation Neutron Source (CSNS), a bench
measurement of a spare ceramic chamber revealed a resonance at 123~kHz, which was identified as the source of a transverse coupled-bunch
instability~\cite{CSNS_EPJP,CSNS_NST}. In the CST simulations at ISIS~\cite{HB23,IPAC26ISIS}, the RF screens were modelled without the magnet yoke, and at CSNS
the yoke was modelled as a perfect electric conductor~\cite{CSNS_NST}.

\begin{figure}[tb]
\centering
\includegraphics[width=0.33\linewidth]{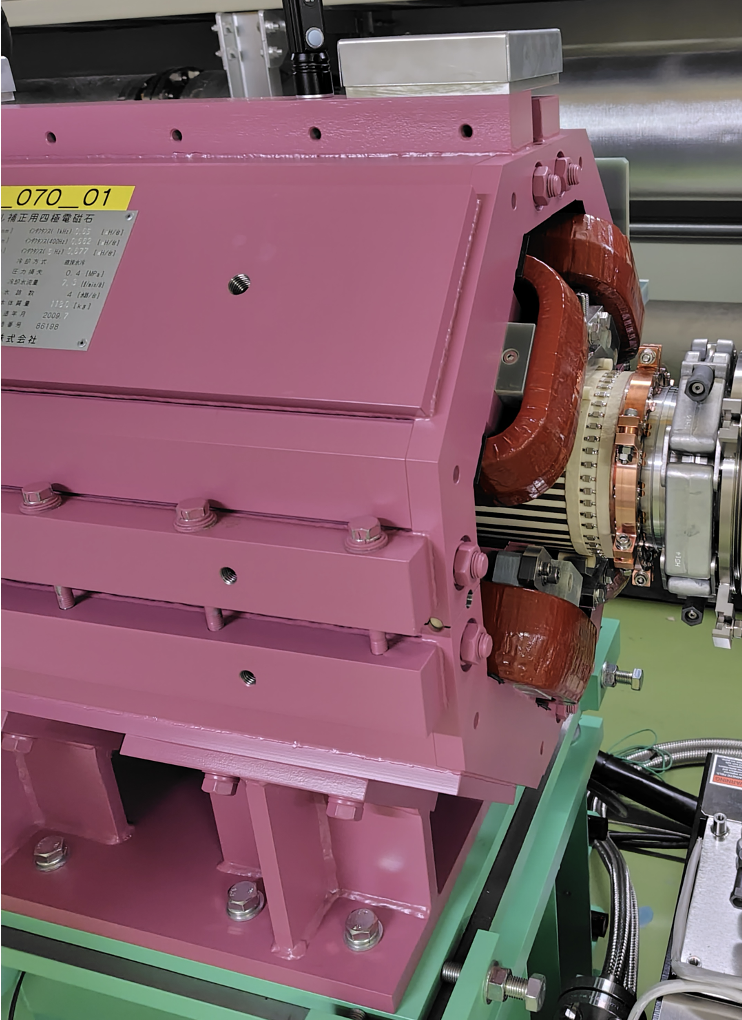}
\caption{RQ magnet with the ceramic chamber, at the capacitor end of the RF shield with the RF-shield clamp.}
\label{fig:photo}
\end{figure}

The J-PARC Main Ring (MR) accelerates a proton beam of eight bunches (harmonic number 9) from 3~GeV to 30~GeV.
In fast extraction (FX), the bunches are extracted in one turn to the Neutrino Experimental Facility. In slow extraction (SX), the beam is
debunched at 30~GeV and extracted over a few seconds by the third-integer resonance to the Hadron Experimental Facility~\cite{Sato25}.
Three spill-feedback quadrupoles (EQ1, EQ2 and RQ) regulate the time structure of the slow-extracted beam. They have ceramic chambers
with copper-strip RF shields closed through 100~nF capacitors at one end~\cite{WEP035}. A photograph of the RQ magnet is shown in Fig.~\ref{fig:photo}. At ISIS the RF screens are placed inside the ceramic
vacuum vessel, whereas in the MR the strips are on the outer surface of the ceramic. Since the ceramic is non-magnetic and displacement currents are
negligible at a few hundred kHz (Sec.~\ref{sec:peec}), this difference does not change the nature of the impedance. There is no spare spill-feedback quadrupole magnet with a ceramic chamber available for bench measurements. To evaluate the impedance, we developed an efficient analysis
method, which combines a
magneto-quasi-static partial element equivalent circuit (PEEC)~\cite{Ruehli74} model of the RF shield with a two-dimensional (2D) field solution
for the laminated magnet yoke.

At a few hundred kHz, a laminated yoke can act as a magnetic material rather than
as a conductor. One purpose of this paper is to show how a laminated yoke changes the resonance frequency $f_r$, the transverse shunt impedance per quality factor $R_\perp/Q$ and the quality factor $Q$. The paper is organized as follows. Section~\ref{sec:structure} describes the ceramic chamber with the RF shield and the spill-feedback quadrupole magnets. Section~\ref{sec:method}
describes the PEEC model, first for the RF shield without the magnet and then with the magnet. Section~\ref{sec:results} gives the results of the PEEC model, and Sec.~\ref{sec:beam} compares them with the resistive-wall impedance of the MR. Section~\ref{sec:discussion} discusses how the material of the RF shield and the thickness of the laminations change the quality factor, and the effect of the excitation of the magnets.

\section{Ceramic chamber, RF shield and magnets}\label{sec:structure}

\subsection{Ceramic chamber and RF shield}\label{sec:chamber}
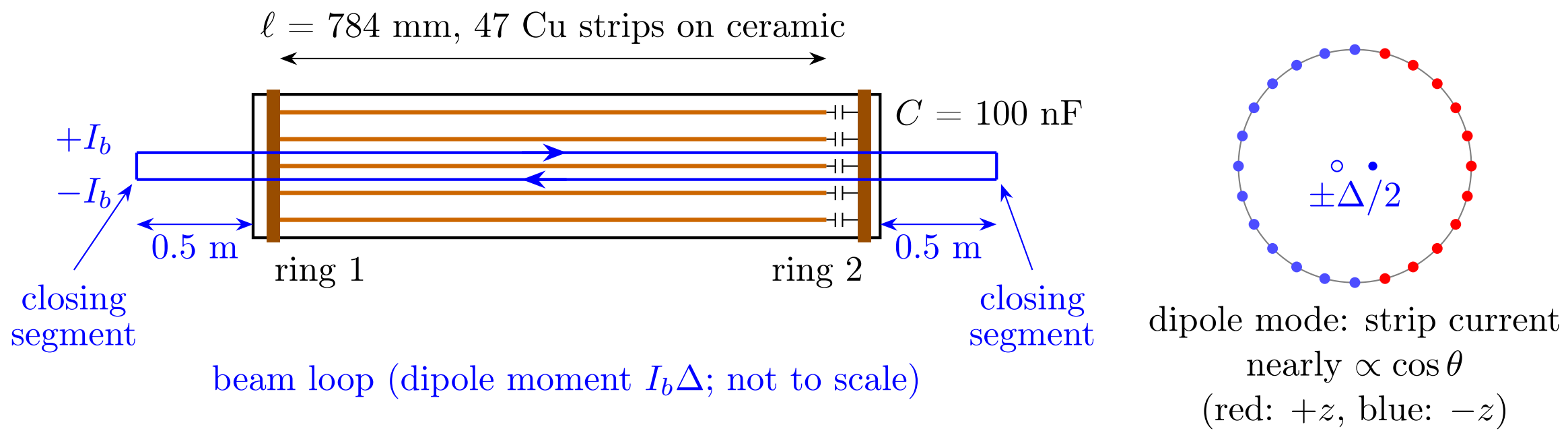
\begin{figure}[tb]
\centering
\begin{tikzpicture}[>=Stealth, scale=0.88]
  \begin{scope}[shift={(0,0)}]
    \draw[thick] (0,0) rectangle (7,1.6);
    \foreach \y in {0.2,0.5,0.8,1.1,1.4} { \draw[orange!80!black, line width=1.2pt] (0.3,\y) -- (6.4,\y); }
    \fill[orange!60!black] (0.15,-0.05) rectangle (0.3,1.65); \node[below] at (0.75,-0.05) {\footnotesize ring 1};
    \fill[orange!60!black] (6.75,-0.05) rectangle (6.9,1.65); \node[below] at (6.3,-0.05) {\footnotesize ring 2};
    \foreach \y in {0.2,0.5,0.8,1.1,1.4} { \draw (6.4,\y) -- (6.5,\y); \draw (6.5,\y-0.08) -- (6.5,\y+0.08); \draw (6.58,\y-0.08) -- (6.58,\y+0.08); \draw (6.58,\y) -- (6.75,\y); }
    \node[anchor=west] at (7.0,1.4) {\footnotesize $C$ = 100 nF};
    \draw[<->] (0.3,2.0) -- (6.4,2.0); \node[above] at (3.35,2.0) {\footnotesize $\ell$ = 784 mm, 47 Cu strips on ceramic};
    \draw[blue, thick] (-1.3,0.95) -- (8.3,0.95);
    \draw[blue, thick] (-1.3,0.65) -- (8.3,0.65);
    \draw[blue, thick] (-1.3,0.65) -- (-1.3,0.95);
    \draw[blue, thick] (8.3,0.65) -- (8.3,0.95);
    \draw[->, blue, thick] (3.0,0.95) -- (3.5,0.95);
    \draw[->, blue, thick] (3.5,0.65) -- (3.0,0.65);
    \node[blue, font=\footnotesize, anchor=east] at (-1.4,1.1) {$+I_b$};
    \node[blue, font=\footnotesize, anchor=east] at (-1.4,0.5) {$-I_b$};
    \draw[blue, <->] (7.0,0.15) -- (8.3,0.15); \node[blue, font=\footnotesize] at (7.65,-0.1) {0.5 m};
    \draw[blue, <->] (-1.3,0.15) -- (0.0,0.15); \node[blue, font=\footnotesize] at (-0.65,-0.1) {0.5 m};
    \node[blue, font=\footnotesize, align=center] (cl1) at (-2.0,-0.9) {closing\\[-2pt]segment}; \draw[blue, ->] (cl1.north) -- (-1.35,0.6);
    \node[blue, font=\footnotesize, align=center] (cl2) at (8.7,-0.9) {closing\\[-2pt]segment}; \draw[blue, ->] (cl2.north) -- (8.35,0.6);
    \node[blue, font=\footnotesize] at (3.5,-1.6) {beam loop (dipole moment $I_b\Delta$; not to scale)};
  \end{scope}
  \begin{scope}[shift={(12.3,0.8)}]
    \draw[gray] (0,0) circle (1.3);
    \foreach \k in {0,...,23} {
      \pgfmathsetmacro{\c}{cos(15*\k)}
      \ifdim \c pt > 0pt \fill[red] ({1.3*cos(15*\k)},{1.3*sin(15*\k)}) circle (0.06); \else \fill[blue!70] ({1.3*cos(15*\k)},{1.3*sin(15*\k)}) circle (0.06); \fi }
    \fill[blue] (0.2,0) circle (0.05); \draw[blue] (-0.2,0) circle (0.06);
    \node[blue] at (0,-0.35) {\footnotesize $\pm\Delta/2$};
    \node[align=center] at (0,-2.25) {\footnotesize dipole mode: strip current\\[-2pt]\footnotesize nearly $\propto\cos\theta$\\[-2pt]\footnotesize (red: $+z$, blue: $-z$)};
  \end{scope}
\end{tikzpicture}
\caption{RF shield (left, side view) and dipole-mode strip current (right, cross-section). Blue: beam loop of two antiparallel wires at $x = \pm\Delta/2$.}
\label{fig:schematic}
\end{figure}

\begin{figure}[tb]
\centering
\includegraphics[width=0.85\linewidth]{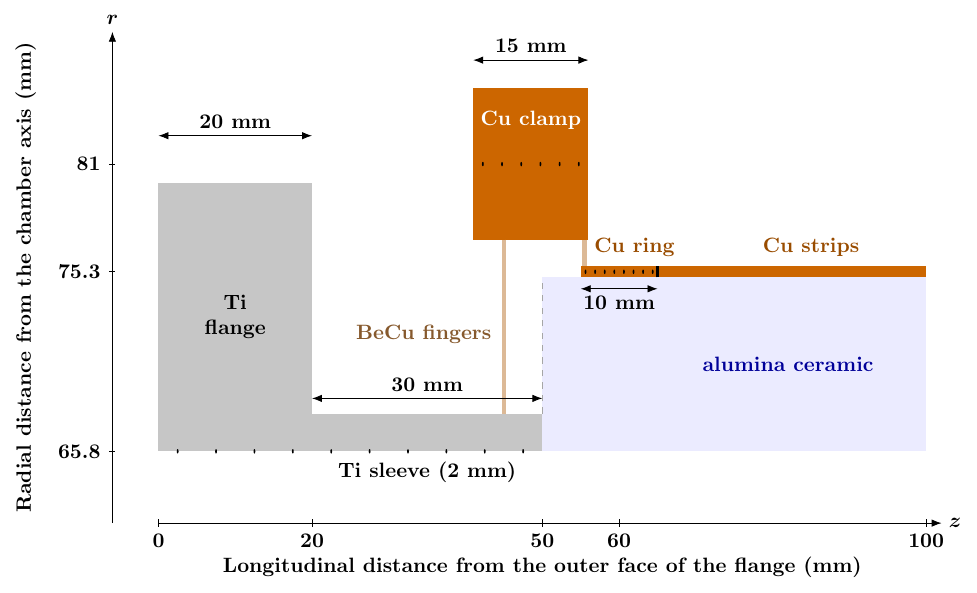}
\caption{Chamber end in the model ($r$--$z$ cross-section, schematic). The dots are the centres of the filaments, and the fingers are represented by radial branches.}
\label{fig:ends}
\end{figure}

The chamber~\cite{WEP035} is 899~mm long, and its central part is an alumina cylinder of 150~mm outer diameter brazed to titanium sleeves at both ends.
The RF shield consists of $N = 47$ copper strips (width 5~mm, thickness 0.3~mm, length $\ell = 784$~mm) fixed on the outer surface of the ceramic
with resin tape at a centre-to-centre spacing of 10.06~mm (radius $a = 75.3$~mm). One end of every strip is connected directly to a copper ring,
and the other end is connected through a 100~nF (1~kV) ceramic capacitor (Murata GRM55DR73A104KW01, X7R) to a second copper ring. Both rings are about
10~mm wide and 0.3~mm thick. The rings are connected to the titanium sleeves by RF-shield clamps at both ends, installed in
2024--2025 to improve the longitudinal impedance~\cite{WEP035}. Each copper clamp (inner diameter 154~mm, width 15~mm) is made of six segments
bolted together, and 36 BeCu fingers (12~mm wide) screwed to the clamp press on the copper ring of the RF shield and on the titanium sleeve. The titanium sleeve
(2~mm thick) connects the ceramic to a titanium flange (outer diameter 160~mm, 20~mm thick). The model used in the analysis is shown in
Figs.~\ref{fig:schematic} and~\ref{fig:ends}.

\subsection{Quadrupole magnets}\label{sec:magnets}
\begin{figure}[tb]
\centering
\includegraphics[width=0.48\linewidth]{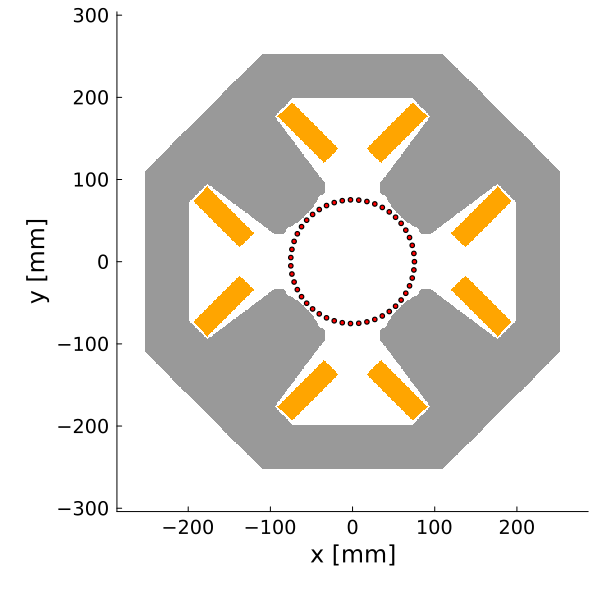}
\caption{Cross-section of the EQ/RQ magnet in the 2D calculation (gray: laminated yoke, orange: coils, red: RF-shield strips).}
\label{fig:xsec}
\end{figure}

EQ1, EQ2 and RQ are normal quadrupoles with a bore (pole-tip) radius $r_0 = 80$~mm and a core length of 0.62~m, made of 0.1~mm thick
laminations of non-oriented electrical steel (ST-100~\cite{ST100}), and clamped by 20~mm thick SUS304 end plates.
The pole profile is hyperbolic ($xy = r_0^2/2$) and spans $\pm18^\circ$ around the pole axis. The pole width increases from about 88~mm at the neck
to 117~mm at the root, the return yoke is 53~mm thick, and the outer contour is an octagon of 506~mm across flats (Fig.~\ref{fig:xsec}).
The pole tips are 4.7~mm away from the RF-shield strips. The coils (EQ: 22 turns per coil, 8.8~mH; RQ: 6 turns per coil, 0.65~mH) lie in the spaces between the poles.
The data sheet of the core material gives a resistivity $\rho = 52~\mu\Omega$cm, a stacking factor of 93\% and an amplitude permeability
of 0.0010--0.0012~H/m at 0.02~T (relative permeability 800--950) between 50~Hz and 20~kHz.
In slow extraction, the EQ current is up to about 100~A, which gives a pole-tip field of up to about 70~mT.

\section{PEEC model}\label{sec:method}

The PEEC model is first applied to the RF shield alone, without the magnet (free space), in Sec.~\ref{sec:peec}. In Sec.~\ref{sec:yoke}, the model is applied to the
RF shield with the magnet, and Sec.~\ref{sec:lam} gives the complex permeability of the laminated magnet yoke.

\subsection{RF shield without the magnet}\label{sec:peec}
The transverse coupling impedance of interest lies in the range of a few hundred kHz, as shown in Sec.~\ref{sec:results}. The corresponding wavelength (about 1~km) is far
longer than the length of the chamber (0.9~m) and the dimensions of the RF-shield elements. Displacement currents and stray capacitances between conductors are therefore neglected
(magneto-quasi-static approximation). The RF shield is divided into straight branches, namely 47 strips along $z$, 47 lumped capacitors and $2\times47$ chords
forming the two rings, as shown in Fig.~\ref{fig:schematic}. Each strip and each chord is further divided into eight parallel filaments across its width, which connect the same two nodes.
The nodes are the points where the branches meet, such as the junctions of each strip with one ring, of each strip with its capacitor and of each capacitor with the other ring. The conductors at each chamber end, namely the copper clamp and the titanium sleeve and flange (hereafter the end conductors), are represented
by further rings of 47 chords that carry the circumferential current. These chords are divided along $z$ into six filaments in the clamp (15~mm) and ten in the titanium sleeve and flange (50~mm). The clamp is
connected to the copper ring and to the titanium sleeve through radial branches representing the fingers, as shown in Fig.~\ref{fig:ends}.
The actual clamp has 36 fingers around the circumference, and the resistance of one finger including the contacts is assumed to be $R_f = 1$~m$\Omega$.
For convenience of modelling, the fingers are represented by 47 branches on each side, the same as the number of strips. The resistance of each branch
is chosen so that the total resistance of the fingers in parallel is unchanged.
The network consists of 2867 branches and 329 nodes in total.

Each filament is treated as a straight line segment carrying a uniform current, and its finite cross section is taken into account below.
The mutual inductance between two segments $i$ and $j$ is given by the Neumann integral
\begin{equation}
M_{ij} = \frac{\mu_0}{4\pi}\int_i\!\int_j \frac{d\mathbf{l}_i\cdot d\mathbf{l}_j}{|\mathbf{r}_i - \mathbf{r}_j|} .
\label{eq:neumann}
\end{equation}
For two parallel $z$-directed segments, Eq.~(\ref{eq:neumann}) can be integrated analytically~\cite{Grover} and written as
\begin{gather}
M_{ij} = \frac{\mu_0}{4\pi}\bigl[F(z_{ib}-z_{ja}) - F(z_{ia}-z_{ja}) - F(z_{ib}-z_{jb}) + F(z_{ia}-z_{jb})\bigr],
\label{eq:grover}\\
F(u) = u\,\sinh^{-1}\frac{u}{d} - \sqrt{u^2+d^2}, \notag
\end{gather}
where $z_{ia}$ and $z_{ib}$ ($z_{ia} < z_{ib}$) are the $z$ coordinates of the two ends $a$ and $b$ of segment $i$, $z_{ja}$ and $z_{jb}$ are those of segment $j$, and $d$ is the transverse distance between them. The self inductance of a segment with a rectangular
cross section of width $w$ and thickness $t$ is given by Eq.~(\ref{eq:grover}) with $d$ replaced by the geometric mean distance (GMD) of the cross section,
approximated by $0.2235(w+t)$~\cite{Rosa}. For the mutual inductance between two different strip filaments, $d$ is replaced by the GMD between their cross sections,
which is evaluated numerically and is approximately the centre-to-centre distance for distant filaments. Since the chords are perpendicular to the strips, the mutual inductances between chords and strips vanish. The mutual inductances between chords, in the same ring and between
the two rings, are computed from Eq.~(\ref{eq:neumann}) by Gauss--Legendre quadrature with 24 points on each chord.
In order to avoid the singularity between adjacent segments, the distance between the two segments $|\mathbf{r}_i - \mathbf{r}_j|$ is replaced by $\sqrt{|\mathbf{r}_i - \mathbf{r}_j|^2 + r_0^2}$
with $r_0$ the mean GMD of the two segments. The resistance of each branch includes the skin effect through an effective
thickness $2\delta(1-e^{-t/2\delta})$ instead of $t$, where $\delta = \sqrt{2\rho/(\omega\mu_0)}$ is the skin depth
($\rho = 1.72\times10^{-8}~\Omega$m for copper and $4.2\times10^{-7}~\Omega$m for titanium, giving, e.g., $\delta \approx 0.15$~mm and 0.73~mm at 200~kHz). This assumes that the branch current is shared equally by the skin layers at the two opposite surfaces. The impedance
$Z_C(\omega)$ of the capacitor branch is obtained from the S-parameter data of the capacitor~\cite{MurataSpar} ($25^\circ$C, no DC bias) and is used directly
in the PEEC model. For example, the data give an equivalent series resistance (ESR) of 74~m$\Omega$ and $C = 102$~nF at 200~kHz.

A rectangular loop of two antiparallel wires is introduced to express the beam dipole moment~\cite{HB23}, as shown in Fig.~\ref{fig:schematic}.
The loop consists of the two $z$-directed wires at $x = \pm\Delta/2$ ($\Delta = 20$~mm) and two short transverse
segments (length $\Delta$) that close it 0.5~m beyond both ends of the chamber. The mutual inductances between the
strips and the $z$-directed wires are also given by Eq.~(\ref{eq:grover}). Since the chords are perpendicular to the $z$-directed wires, only their small mutual
inductances with the two transverse segments remain (see Fig.~\ref{fig:schematic}), and these are computed from Eq.~(\ref{eq:neumann}).
The circuit is formulated by modified nodal analysis~\cite{Ho75} as
\begin{equation}
\begin{pmatrix}\mathsf{Z}(\omega) & -\mathsf{A}^T\\ \mathsf{A} & 0\end{pmatrix}
\begin{pmatrix}\mathbf{I}\\ \boldsymbol{\phi}\end{pmatrix}
= \begin{pmatrix}-j\omega\,\mathbf{M}^\mathrm{beam} I_b\\ 0\end{pmatrix},
\label{eq:mna}
\end{equation}
where $\mathbf{I}$ and $\boldsymbol{\phi}$ are the branch currents and the node potentials. $\mathsf{Z}(\omega) = j\omega\mathsf{L} + \mathsf{D}(\omega)$ is the
$2867\times2867$ branch impedance matrix, where $\mathsf{D}$ is a diagonal matrix whose elements are the resistances $R_b$ of the strips, chords and
fingers and the impedances $Z_C$ of the capacitors. The node--branch incidence matrix $\mathsf{A}$ describes how the branches are connected. Its element $A_{nb}$ is $+1$
if branch $b$ starts at node $n$, $-1$ if it ends at node $n$, and 0 otherwise (one node is grounded and its row is omitted). With this matrix, the first block row
of Eq.~(\ref{eq:mna}) states that the voltage across each branch equals the potential difference between its end nodes, and the second block row is
Kirchhoff's current law at each node. $\mathbf{M}^\mathrm{beam}$ is the vector of mutual inductances between the branches and the beam loop, and $I_b$ is the current of the beam loop. Since all currents and the induced voltage are proportional to $I_b$ (the circuit is linear), the transverse coupling impedance does not depend on $I_b$, and we set $I_b = 1$~A.
At each frequency, Eq.~(\ref{eq:mna}) is a dense complex linear system for the $2867 + 328 = 3195$ unknowns in $\mathbf{I}$ and $\boldsymbol{\phi}$, which is solved
by LU decomposition using LAPACK and BLAS through the linear-algebra library of Julia. The computation takes about 1.3~s per frequency on a laptop computer.

The beam loop sees an impedance change
\begin{equation}
\Delta Z = \frac{j\omega\sum_b M_b^\mathrm{beam}[\mathbf{I}]_b}{I_b} ,
\label{eq:dZ}
\end{equation}
where $[\mathbf{I}]_b$ and $M_b^\mathrm{beam}$ are the $b$th elements of $\mathbf{I}$ and $\mathbf{M}^\mathrm{beam}$, namely the current in branch $b$ and its mutual inductance with the beam loop.
The longitudinal coupling impedance between a source at $x_s$ and a test particle at $x_t$ is expanded as
\begin{equation}
Z_\parallel(x_s,x_t) = Z_\parallel^{(0)} + \frac{\omega}{c}\,Z_\perp\,x_s x_t + \dots ,
\label{eq:PW}
\end{equation}
where the coefficient of the dipole term is the transverse coupling impedance $Z_\perp$ (Pan\-of\-sky--Wenzel theorem~\cite{PW56}). When Eq.~(\ref{eq:PW}) is summed over
the four source--test combinations of the two wires in $\Delta Z$, the monopole term cancels and the dipole term remains. The impedance change is then written as
\begin{equation}
\Delta Z = \sum_{s,t=\pm}\sigma_s\sigma_t Z_\parallel(x_s,x_t) = \frac{\omega}{c}Z_\perp\Delta^2 ,
\label{eq:foursum}
\end{equation}
where $\sigma_\pm = \pm1$ is the sign of the current in the wire at $x = \pm\Delta/2$. The transverse coupling impedance is therefore obtained from $\Delta Z$ in Eq.~(\ref{eq:dZ}) as
\begin{equation}
Z_\perp = \frac{c\,\Delta Z}{\omega\Delta^2},
\label{eq:twowire}
\end{equation}
which is the usual two-wire formula~\cite{Nassibian79}.

\begin{figure}[tb]
\centering
\includegraphics[width=0.6\linewidth]{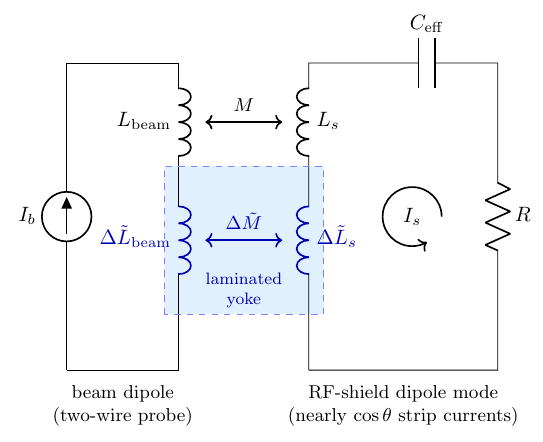}
\caption{Single-loop equivalent circuit of the beam coupling to the dipole mode of the RF shield (schematic). Blue: complex inductances added by the laminated yoke.}
\label{fig:circuit}
\end{figure}

The following single-loop picture is not used in the calculation but helps to interpret the results. When the RF shield is approximated by a single loop of inductance $L_s$, effective capacitance $C_\mathrm{eff}$ and resistance $R$, coupled to the beam loop by the
mutual inductance $M$, the impedance change becomes the reflected impedance of a transformer,
\begin{equation}
\Delta Z = \frac{\omega^2M^2}{Z_2},\qquad Z_2 = R + j\Bigl(\omega L_s - \frac{1}{\omega C_\mathrm{eff}}\Bigr) ,
\label{eq:singleloop}
\end{equation}
where $Z_2$ is the series impedance of the RF-shield loop (the secondary loop on the right in Fig.~\ref{fig:circuit}).
The real part of $Z_\perp$ peaks at the series resonance of the loop, $\omega_r = 1/\sqrt{L_s C_\mathrm{eff}}$, where $Z_2 = R$. We define the shunt impedance
$R_\perp$ as the peak value of $\mathrm{Re}\,Z_\perp$ and the quality factor as $Q = f_r/\Delta f$, where $\Delta f$ is the full width at
half maximum. In the single-loop picture, $Q$ can be written as $\omega_r L_s/R$. From Eqs.~(\ref{eq:twowire}) and (\ref{eq:singleloop}), $R_\perp = c\,\omega_r M^2/(R\Delta^2)$, and hence
\begin{equation}
\frac{R_\perp}{Q} = \frac{c\,M^2}{L_s\,\Delta^2}
\label{eq:RQ}
\end{equation}
depends only on geometry, while the losses affect only $Q$. Far above the resonance ($\omega \gg \omega_r$), $Z_2 \to j\omega L_s$ and
$Z_\perp \to -j\,cM^2/(L_s\Delta^2) = -jR_\perp/Q$, a frequency-independent reactance. In this limit, the capacitors are effectively short-circuited, and the RF shield behaves
as a closed conducting cage.
Figure~\ref{fig:circuit} shows this single-loop equivalent circuit, which is a schematic of what the PEEC model solves with the full network.
The beam (or the two-wire probe) is the primary, the dipole mode of the RF shield (strip currents nearly $\propto\cos\theta$) is the secondary with the loop
inductance $L_s$, the effective capacitance $C_\mathrm{eff}$ and the resistance $R$ of the strips, rings and capacitors, and the two are
coupled by the mutual inductance $M$, which exists even without the magnet around the RF shield. The elements drawn in blue represent the laminated yoke of the magnet and are explained in
Sec.~\ref{sec:yoke}.

\subsection{RF shield with the magnet}\label{sec:yoke}
With the magnet, the laminated yoke surrounds the RF shield and changes the magnetic flux linked by each loop of Fig.~\ref{fig:circuit}. The inductances of the
RF-shield loop ($L_s$) and of the beam loop ($L_\mathrm{beam}$) and the mutual inductance ($M$) between them are modified, and the magnetic loss in the laminations
increases the resistances of the loops. In Fig.~\ref{fig:circuit}, these effects are drawn as an additional coupling $\Delta\tilde M$ and the inductances $\Delta\tilde L_s$ and $\Delta\tilde L_\mathrm{beam}$
(due to the beam seeing the iron directly) in series with the RF-shield loop and the beam loop, respectively. Since the permeability of the iron is
complex, $\mu' - j\mu''$, these inductance changes are complex. Their real parts (from $\mu'$) increase the reactances of the loops, and their
imaginary parts (from $\mu''$) represent the magnetic loss, which is dissipated in the iron but appears as the resistive parts of the complex inductances drawn in blue in Fig.~\ref{fig:circuit}.
Since the magnet core length of 0.62~m is large compared with the bore radius, we obtain these changes per unit length approximately from a two-dimensional (2D)
magneto-quasi-static solution of Maxwell's equations in the magnet cross-section.

Since all currents (in the strip filaments and the beam wires) flow along $z$ in 2D, the magnetic flux density $\mathbf{B}$ lies in the cross-section and is
described by the $z$-component $A_z(x,y)$ of the vector potential as $\mathbf{B} = \nabla\times(A_z\hat{\mathbf{z}})$. Ampère's law $\nabla\times(\nu\mathbf{B}) = \mathbf{J}$
then gives
\begin{equation}
-\nabla\cdot(\nu\nabla A_z) = J_z ,
\label{eq:Az}
\end{equation}
where $J_z$ is the $z$-directed current density of the strip filaments and the beam wires, and $\nu$ is the reluctivity, which is $1/(\mu_0\mu_\mathrm{iron})$ in the iron with the complex relative permeability $\mu_\mathrm{iron}$ of the laminations (Sec.~\ref{sec:lam}).
For a current $I_j$ in filament $j$ alone, the difference of $A_z$ between filament $i$ and the outer boundary at $r_b$ is the magnetic flux per unit
length, which gives the mutual inductance per unit length
\begin{equation}
L'_{ij} = \frac{A_z^{(j)}(\mathbf{r}_i) - A_z^{(j)}(r_b)}{I_j} = \frac{A_z^{(j)}(\mathbf{r}_i)}{I_j},
\label{eq:Lij}
\end{equation}
where $A_z^{(j)}$ is the solution of Eq.~(\ref{eq:Az}) for this current and $\mathbf{r}_i$ is the position of filament $i$.
Since the dipole mode of the strips and the beam loop both carry zero net current, their $A_z$ decays as $1/r$ at large distances. The boundary condition $A_z = 0$ is
imposed at $r_b = 1.5$~m, which is chosen as a distance at which $A_z$ is sufficiently small, far outside the yoke.
Equation~(\ref{eq:Az}) is solved on a polar
finite-volume grid with 325 cells in the radial direction and 1504 cells in the azimuthal direction. The eight filaments of each strip are placed at the centres of every other cell in the azimuthal direction,
and the two beam wires are also placed at cell centres.
Equation~(\ref{eq:Az}) is solved separately with (yoke) and without (air) the yoke on the same grid, and then their differences give the changes of the per-unit-length inductances caused by the yoke,
\begin{equation}
\begin{aligned}
\Delta L'_{ij} &= L'_{ij}\big|_\mathrm{yoke} - L'_{ij}\big|_\mathrm{air},\\
\Delta M'_i &= M'_i\big|_\mathrm{yoke} - M'_i\big|_\mathrm{air},\\
\Delta L'_\mathrm{beam} &= L'_\mathrm{beam}\big|_\mathrm{yoke} - L'_\mathrm{beam}\big|_\mathrm{air},
\end{aligned}
\label{eq:dL}
\end{equation}
where $\Delta L'_{ij}$ is between strips $i$ and $j$, $\Delta M'_i$ is between strip $i$ and the beam loop, and $\Delta L'_\mathrm{beam}$ is the self-inductance of the beam
loop. For the two beam wires ($\pm I_b$ at $\pm\Delta/2$), $M'_i$ is $A_z$ at strip $i$ divided by $I_b$, and $L'_\mathrm{beam}$ is the
difference of $A_z$ between the two beam wires divided by $I_b$. The differences of Eq.~(\ref{eq:dL}) nearly cancel the dependence on the cell size and the error of the outer boundary
condition $A_z = 0$ at $r_b$. The 2D solver is verified by the following simple model. For a $\cos\theta$ surface current on a continuous cylinder with the RF-shield radius $a = 75.3$~mm, surrounded by a thick iron cylinder ($\mu_r = 1000$, inner radius $r_\mathrm{in} = 80$~mm), the image method analytically gives an increase of the dipole-mode inductances by the fraction $\kappa(a/r_\mathrm{in})^2 = 0.88$, where $\kappa = (\mu_r-1)/(\mu_r+1)$. The solver reproduces it to within 0.7\% for strip currents proportional to $\cos\theta$.
The changes are multiplied by the core length $\ell_m = 0.62$~m and added to the PEEC matrices as
\begin{equation}
\mathsf{Z}(\omega) = j\omega\bigl(\mathsf{L} + \ell_m\Delta\mathsf{L}'(\omega)\bigr) + \mathsf{D}(\omega),\qquad
\mathbf{M}^\mathrm{beam} \to \mathbf{M}^\mathrm{beam} + \ell_m\Delta\mathbf{M}'(\omega),
\label{eq:Zyoke}
\end{equation}
where $\Delta\mathsf{L}'(\omega)$ and $\Delta\mathbf{M}'(\omega)$ are the matrix of $\Delta L'_{ij}$ and the vector of $\Delta M'_i$ in Eq.~(\ref{eq:dL}), and $j\omega\ell_m\Delta L'_\mathrm{beam}$ is added to $\Delta Z$ (due to the beam seeing the iron directly). Since $\mu_\mathrm{iron}$ depends on frequency, the changes are computed at 31 frequencies between 10~kHz and 1~MHz and interpolated in $\log f$.
The free-space partial inductance matrix $\mathsf{L}$ depends only on
geometry and is real, and the losses of the free-space circuit are only in the diagonal matrix $\mathsf{D}$. The yoke terms are complex because
$\mu_\mathrm{iron} = \mu' - j\mu''$ is complex, and $\mathsf{Z}$ becomes complex symmetric. The real parts of $\ell_m\Delta\mathsf{L}'$ and $\ell_m\Delta\mathbf{M}'$ increase the reactances, while
$j\omega$ times their imaginary parts is real, i.e., resistive, and represents the magnetic loss in the iron. The resistive part of $\ell_m\Delta\mathsf{L}'$ appears not only on the diagonal but
also off the diagonal, as mutual resistances between different strips, because the loss in the iron is caused by the superposition of the fields of all
strips. The complex inductances $\Delta\tilde L_s$, $\Delta\tilde M$ and $\Delta\tilde L_\mathrm{beam}$ in Fig.~\ref{fig:circuit} are the dipole-mode parts of $\ell_m\Delta L'_{ij}$, $\ell_m\Delta M'_i$ and $\ell_m\Delta L'_\mathrm{beam}$,
and the mutual resistances between strips are included in the imaginary part of $\Delta\tilde L_s$.

At the two ends of the yoke, the end faces are covered by the 20~mm SUS304 end plates. The magnetic field of the strip and beam currents outside the plates is parallel to the plates and penetrates them only within the skin depth (0.96~mm at 200~kHz), which is much smaller than the plate thickness. The field therefore does not reach the iron, and the loss occurs on the plate surface with the surface impedance $Z_s = (1+j)R_s$, where $R_s = \rho_\mathrm{SUS}/\delta_\mathrm{SUS} = 7.5\times10^{-4}~\Omega$ at 200~kHz. This
loss is estimated as $P = \frac12\mathrm{Re}\,Z_s\int|H_t|^2dS$, where $H_t$ is the tangential component of the magnetic field on the plate surfaces. This field is approximated by the free-space field of the strip and beam currents in 2D. Since $H_t$ is the sum of the fields of all strips and the beam, the loss is a quadratic form of their currents. The end plates therefore add impedances to all elements of $\mathsf{Z}(\omega)$ in Eq.~(\ref{eq:Zyoke}). In the same way as the yoke terms, they also add mutual impedances between the strips and the beam loop to $j\omega\mathbf{M}^\mathrm{beam}$, and a self-impedance of the beam loop to $\Delta Z$.

Since the flux of the RF shield returns through the iron, the flux in the spaces between the poles where the coils lie
(Fig.~\ref{fig:xsec}) is weak, and the loss in the coils can be neglected. The coils are therefore not included in the model for the main results below. However, the effect of the weak flux excluded by the coils, treated as perfect conductors, is estimated in Sec.~\ref{sec:resyoke}.

\subsection{Complex permeability of the laminated yoke}\label{sec:lam}
The complex relative permeability $\mu_\mathrm{iron}$ of the laminations in Eq.~(\ref{eq:Az}) is obtained as follows. Since the laminations are insulated from each other along $z$, no eddy
current flows along $z$ in 2D. The eddy currents circulate within each lamination across its thickness $d$, and their effect is represented by the classical complex
permeability of a lamination with in-plane flux~\cite{Lammeraner},
\begin{gather}
\mu_\mathrm{lam} = \mu_r\,\frac{\tanh(kd/2)}{kd/2},\qquad \mu_\mathrm{iron} = F\mu_\mathrm{lam} + (1-F),
\label{eq:mulam}\\
k = \frac{1+j}{\delta},\qquad \delta = \sqrt{\frac{2\rho}{\omega\mu_0\mu_r}}, \notag
\end{gather}
where $\delta$ is the skin depth in the steel and $F$ is the stacking factor.

\begin{table}[htb]
\caption{Data-sheet iron loss of ST-100 (0.1~mm) compared with the classical eddy-current loss of the lamination model.}
\label{tab:excess}
\centering
\small\setlength{\tabcolsep}{4pt}
\begin{tabular}{lcccc}
\toprule
 & $W_{5/2000}$ & $W_{2/5000}$ & $W_{1/10000}$ & $W_{0.5/20000}$ \\
Peak flux density $B$ [T] & 0.5 & 0.2 & 0.1 & 0.05 \\
Frequency $f$ [kHz] & 2 & 5 & 10 & 20 \\
\midrule
(a) Iron loss, data sheet [W/kg] & 25.3 & 17.0 & 12.9 & 10.2 \\
(b) Classical eddy-current loss, Eq.~(\ref{eq:pdens}) [W/kg] & 4.13 & 4.13 & 4.13 & 4.12 \\
Ratio (a)/(b) & 6.1 & 4.1 & 3.1 & 2.5 \\
\bottomrule
\end{tabular}

\end{table}

The ST-100 data sheet~\cite{ST100} gives the iron loss per unit mass at constant $f\!\cdot\!B = 1000$~T\,Hz, where $B$ is the peak flux density
((a) in Table~\ref{tab:excess}). The classical eddy-current loss obtained from Eq.~(\ref{eq:mulam}) is listed as (b) in
Table~\ref{tab:excess}. It is calculated from the loss density
\begin{equation}
p = \frac{1}{2\mu_0}\,\omega B^2\,\mathrm{Im}\bigl(1/\mu_\mathrm{lam}\bigr),
\label{eq:pdens}
\end{equation}
which is derived from $p = \frac{1}{2\mu_0}\mathrm{Re}\bigl(j\omega BB^*/\mu_\mathrm{lam}^*\bigr)$. It is converted to the loss per unit mass with an
assumed steel density of 7650~kg/m$^3$. The classical eddy-current loss (b) is nearly constant at about 4.1~W/kg for constant $f\!\cdot\!B$.
This shows that, at the frequencies of the data sheet (2--20~kHz), the 0.1~mm laminations are in the low-frequency regime,
where Eq.~(\ref{eq:pdens}) can be approximated as $p = \pi^2f^2B^2d^2/(6\rho)$, which depends only on $f\!\cdot\!B$ and is independent of $\mu_r$. The measured
losses (a) are 2.5--6.1 times the classical values (b), and the ratio decreases with frequency. The difference is mainly due to the excess (anomalous) loss, which arises from eddy currents concentrated near the moving domain walls, and the hysteresis loss~\cite{Bertotti88}. The conditions of interest here ($\sim$200~kHz and the very small flux density induced by the beam) are outside the range of the data sheet. The ratio $k_\mathrm{exc}$, defined as (a)/(b) in Table~\ref{tab:excess}, decreases with frequency approximately as $f^{-0.4}$, and its extrapolation
along constant $f\!\cdot\!B$ approaches 1 at $\sim$200~kHz. At the very small flux density induced by the beam, the hysteresis loss,
which scales as $B^3$ in the Rayleigh region (the low-field region where the permeability increases linearly with the field), becomes negligible. However, the excess loss does not vanish at small amplitude, since the domain walls still move by the so-called reversible bowing and induce local eddy currents. This part cannot be estimated from the available data. We therefore
calculate two cases, $k_\mathrm{exc} = 1$ (classical eddy current only) and $k_\mathrm{exc} = 2.5$ (the
ratio at 20~kHz). From Eq.~(\ref{eq:pdens}), the loss at a given $B$ is proportional to $\mathrm{Im}(1/\mu_\mathrm{lam})$, and $\mathrm{Im}(1/\mu_\mathrm{lam})$
is multiplied by $k_\mathrm{exc}$ while its real part, which determines the inductive response, is kept. The resultant $\mu_\mathrm{lam}$ at 20~kHz for $\mu_r = 1000$
is $930 - 231j$ ($k_\mathrm{exc} = 1$) and $712 - 443j$ ($k_\mathrm{exc} = 2.5$), whose magnitudes $|\mu_\mathrm{lam}| = 958$ and 838 are close to the
data-sheet amplitude permeability of about 900 at 0.02~T and 20~kHz (read from the curve).

The calculations of the transverse coupling impedance with the magnet in Sec.~\ref{sec:resyoke} use the data-sheet values $\rho = 52~\mu\Omega$cm and $F = 0.93$
(Sec.~\ref{sec:magnets}). The frequency dependence of $\mu_\mathrm{lam}$ caused by the eddy currents is included through Eq.~(\ref{eq:mulam}), while the
intrinsic permeability $\mu_r$ and the factor $k_\mathrm{exc}$ are taken as constant in frequency. The data sheet gives the iron properties only up to 20~kHz.
Above the frequency range of the data sheet, $\mu_r$ may be lower since the domain-wall motion cannot follow the field, and
$k_\mathrm{exc}$ may also depend on frequency. Since the full width at half maximum $\Delta f$ is less than 4\% of the resonance frequency
(Sec.~\ref{sec:resyoke}), the frequency dependence of $\mu_r$ and $k_\mathrm{exc}$ around the resonance can be neglected. We take two values of $\mu_r$, 1000, close to the data-sheet amplitude permeability, and 500, allowing for the lower permeability at small
amplitude and above 20~kHz. Each is calculated with $k_\mathrm{exc} = 1$ and 2.5. For example, at 200~kHz, the frequency of interest,
and $\mu_r = 1000$, the skin depth $\delta$ is $\sim$26~$\mu$m, smaller than half the lamination thickness, and $\mu_\mathrm{lam}$ is $\sim$$257 - 272j$,
which gives a loss angle close to $45^\circ$.

\section{Results of the PEEC model}\label{sec:results}

\subsection{Resonance without the magnet}\label{sec:resfree}
\begin{figure}[tb]
\centering
\includegraphics[width=0.85\linewidth]{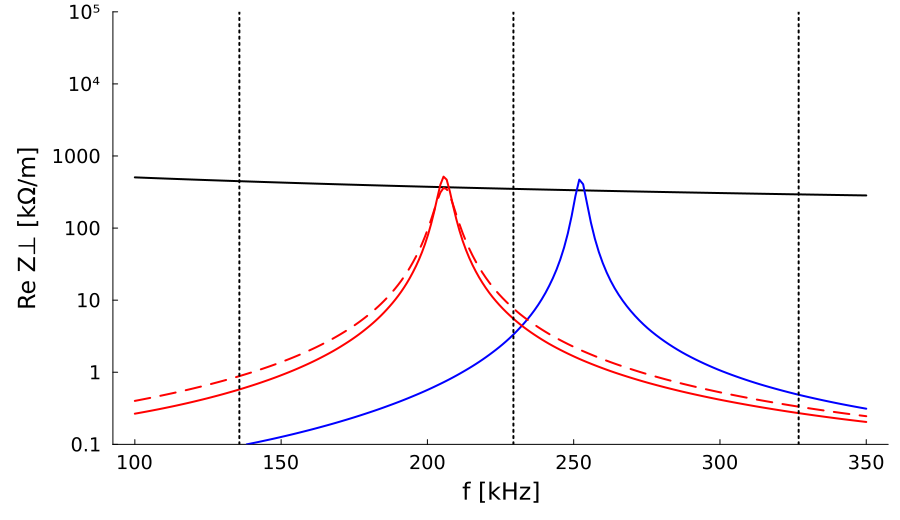}
\caption{$\mathrm{Re}\,Z_\perp$ of one chamber with the magnet (red, $\mu_r = 1000$, $k_\mathrm{exc} = 1$ solid and 2.5 dashed) and without the magnet (blue), and the resistive wall of the whole ring (black). Dotted lines: unstable sidebands in slow extraction.}
\label{fig:ReZ}
\end{figure}

We first present the results for the RF shield without the magnet (free space). The transverse impedance $Z_\perp(f)$ is obtained by the method described in Sec.~\ref{sec:peec}. At each frequency, the capacitor impedance $Z_C(\omega)$ is interpolated from the
S-parameter data, so that the frequency dependence of $Z_C$ is included in the
resonance curve. When $f_r$ and $R_\perp$ are defined as the frequency and the value at the maximum of $\mathrm{Re}\,Z_\perp(f)$, the PEEC model in free space gives $f_r = 252.3$~kHz and $R_\perp = 0.49$~M$\Omega$/m per chamber,
as shown by the blue curve in Fig.~\ref{fig:ReZ}. The resonance curve gives $Q = 63$, and hence $R_\perp/Q = 7.76$~k$\Omega$/m. These values are obtained under the following conditions. Each strip and each chord is divided into eight filaments, the clamp into six and the titanium sleeve and flange into ten, the finger resistance is $R_f = 1$~m$\Omega$, and the beam loop with $\Delta = 20$~mm is closed 0.5~m beyond the chamber ends. The dependence on these conditions is shown later.
The strip current distribution computed by the PEEC model at $f_r$ agrees with $\cos\theta$ within an rms deviation of 1.7\% of its amplitude,
which shows that the resonance is the dipole
mode.

In the real strips, the current distribution across the strip width is not uniform.
Each strip is therefore divided into eight filaments, as described above.
If the number of filaments per strip is changed from eight to one, $f_r$ changes by $-0.7\%$ and $R_\perp/Q$ by $+0.6\%$, but $Q$ increases by 16\%.
For a given net current of a strip, the loss is smallest when the currents in its filaments are equal, as with a single filament.
With eight filaments, the filament currents are determined mainly by the inductances at this frequency and are not equal, which increases the loss of the strips.
Since the current distribution across the strip width approaches the real one as the number of filaments is increased, an extrapolation from the results with eight and 16 filaments suggests that the real $Q$ is about 6\% lower than the value with eight filaments, while $f_r$ and $R_\perp/Q$ do not change.

Each chord of the ring is divided into eight filaments,
since its length (10~mm) is comparable to its width (the ring width of 10~mm). Even if the number of filaments per chord is changed from eight to one or 16, $f_r$, $R_\perp/Q$ and $Q$ change by less than 0.2\%.

The resonance parameters $f_r$, $R_\perp$ and $Q$ are insensitive to the separation $\Delta$ of the beam loop and to its extension beyond the chamber ends (Fig.~\ref{fig:schematic}).
If $\Delta$ is changed from 20 to 10~mm or the extension from 0.5 to 1~m, they change by less than 1\%.

If the end conductors mentioned in Sec.~\ref{sec:peec} are removed, the PEEC model gives $f_r = 239.9$~kHz, $R_\perp/Q = 6.96$~k$\Omega$/m and $Q = 63$. The copper clamps raise $f_r$ by 2.3\%, and
the titanium sleeves and flanges raise it by a further 2.8\%, since they carry part of the circumferential current of the rings and reduce the loop
inductance $L_s$. The end conductors also increase $R_\perp/Q$ by 12\%, while $Q$ hardly changes. The increase of $R_\perp/Q$ is explained by the reduction of $L_s$,
since $R_\perp/Q$ is proportional to $M^2/L_s$ in the single-loop picture (Eq.~(\ref{eq:RQ})) and the end conductors, which are perpendicular to the
$z$-directed wires of the beam loop, hardly change $M$. If the finger resistance is changed from 1~m$\Omega$ to
0.3 or 3~m$\Omega$, $Q$ changes by +2\% or $-5\%$ without affecting $f_r$ and $R_\perp/Q$. If the number of filaments in the end conductors is doubled, $Q$ changes
by $-1.4\%$.

The PEEC results are compared with simple analytic estimates. In the analytic single-loop picture, the current $I_s$ of the RF-shield loop on the right of Fig.~\ref{fig:circuit} is the total current in the strips
on one side shown in red in Fig.~\ref{fig:schematic} (right).
The inductance $L_s$ and the effective capacitance $C_\mathrm{eff}$ in Fig.~\ref{fig:circuit} can be expressed as $L_s = (\mu_0\pi/8)\ell$ and $C_\mathrm{eff} = 2NC/\pi^2$, by requiring that the loop with $I_s$ stores the same magnetic energy as
a $\cos\theta$ surface current on a continuous cylinder replacing the strips (neglecting the ends) and the same capacitor energy as the $N$ capacitors with currents
proportional to $\cos\theta$. With $\ell = 784$~mm, $N = 47$ and the nominal capacitance $C = 100$~nF, $L_s$ and $C_\mathrm{eff}$ are 0.39~$\mu$H and 0.95~$\mu$F, which give
$f_s = 1/(2\pi\sqrt{L_sC_\mathrm{eff}}) = 262$~kHz, 4\% higher than the PEEC value (252.3~kHz). The analytic estimate neglects the rings and the end conductors.
The inductance of the rings lowers $f_r$, and the end conductors raise it by carrying part of the circumferential current of the rings.
In the single-loop picture (Sec.~\ref{sec:peec}), both $R_\perp/Q$ and $|\mathrm{Im}\,Z_\perp|$ far above the resonance are equal to
$cM^2/(L_s\Delta^2)$ of Eq.~(\ref{eq:RQ}), which depends only on the geometry. The PEEC model gives $\mathrm{Im}\,Z_\perp = -7.88$~k$\Omega$/m at 5~MHz,
which is about 20 times $f_r$ and is taken as the high-frequency limit. This value agrees within 2\% with $-R_\perp/Q = -7.76$~k$\Omega$/m, obtained from the peak
value and the half-maximum width of the $\mathrm{Re}\,Z_\perp$ resonance curve. This small difference shows that the PEEC result is consistent with the single-loop picture of
Eq.~(\ref{eq:RQ}). With $L_s$ above and $M = \mu_0\Delta\ell/(4a)$, which is derived from the uniform field $\mu_0 I_s/(4a)$ of the same $\cos\theta$ surface current
threading the beam loop, $R_\perp/Q = cM^2/(L_s\Delta^2)$ of Eq.~(\ref{eq:RQ}) becomes $Z_0\ell/(2\pi a^2) = 8.3$~k$\Omega$/m, where $Z_0 = \mu_0c = 376.7~\Omega$
is the impedance of free space. Far above the resonance, $\mathrm{Im}\,Z_\perp$ therefore approaches $-Z_0\ell/(2\pi a^2)$, which is the image-current
impedance of a perfectly conducting cylinder. The PEEC value of $|\mathrm{Im}\,Z_\perp|$ at 5~MHz is 5\% smaller than $Z_0\ell/(2\pi a^2)$.
This discrepancy is reduced as the RF shield is made longer and approaches the condition of the single-loop picture assuming infinite length.

\subsection{Resonance with the magnet}\label{sec:resyoke}
\begin{table}[tb]
\caption{Resonance parameters of one chamber with the magnet. $Q$ is given without (yoke only) and with (end plates) the loss on the SUS304 end plates, and $f_r$ and $R_\perp/Q$ include it.}
\label{tab:results}

\centering
\small\setlength{\tabcolsep}{4pt}
\begin{tabular}{lcccc}
\toprule
 & $f_r$ & $R_\perp/Q$ & $Q$ & $Q$ \\
 & [kHz] & [k$\Omega$/m] & yoke only & end plates \\
\midrule
$\mu_r = 1000$, $k_\mathrm{exc} = 1$ & 205.77 & 12.21 & 43.1 & 42.9 \\
$\mu_r = 1000$, $k_\mathrm{exc} = 2.5$ & 205.85 & 12.17 & 30.0 & 29.9 \\
$\mu_r = 500$, $k_\mathrm{exc} = 1$ & 206.04 & 12.17 & 40.9 & 40.7 \\
$\mu_r = 500$, $k_\mathrm{exc} = 2.5$ & 206.14 & 12.13 & 27.4 & 27.3 \\
lossless iron ($\mu_r = 1000$, $\mathrm{Im}\,1/\mu = 0$) & 205.77 & 12.22 & 61.4 & 60.8 \\
\bottomrule
\end{tabular}

\end{table}

The transverse impedance $Z_\perp(f)$ with the magnet is obtained by the method described in Secs.~\ref{sec:peec} and \ref{sec:yoke}. The values of $f_r$, $R_\perp$ and
$Q$ are obtained as in Sec.~\ref{sec:resfree}. Table~\ref{tab:results} summarizes the results for the four material cases of Sec.~\ref{sec:lam}.
The last row is a reference case of lossless iron, in which $\mu_r = 1000$ and the imaginary part of
$1/\mu_\mathrm{lam}$ is set to zero ($k_\mathrm{exc} = 0$). Since the yoke in this case changes the inductances but causes no loss, $Q$ is determined
only by the losses of the RF-shield circuit consisting of strips, rings, capacitors and end conductors. It is used below to separate the yoke loss from the losses of the RF-shield circuit.
The obtained resonance curves for $\mu_r = 1000$ are plotted in red in Fig.~\ref{fig:ReZ}, with solid and dashed lines for $k_\mathrm{exc} = 1$ and 2.5, respectively. Those for $\mu_r = 500$ with the same $k_\mathrm{exc}$ nearly coincide with them and are not shown.
The resonance frequency and $R_\perp/Q$ are about 206~kHz and 12~k$\Omega$/m, which hardly depend on the material parameters
$\mu_r$ and $k_\mathrm{exc}$. The quality factor $Q$ depends mainly on $k_\mathrm{exc}$ and only weakly on $\mu_r$.
The full width at half maximum $\Delta f$ is 5--8~kHz. Compared with the results without the magnet (Sec.~\ref{sec:resfree}), $f_r$ is 18\% lower and $R_\perp/Q$ is 57\% higher.
In the single-loop picture, the yoke increases both the loop inductance $L_s$ and the coupling $M$ to the beam. The increase of $L_s$ lowers $f_r$. The increase of $R_\perp/Q$ indicates that $M^2$ increases more than $L_s$, since $R_\perp/Q$
is proportional to $M^2/L_s$ (Eq.~(\ref{eq:RQ})).

The column ``end plates'' of Table~\ref{tab:results} and the resonance curves in Fig.~\ref{fig:ReZ} include the loss on the SUS304 end plates (Sec.~\ref{sec:yoke}).
If this loss is neglected, $Q$ increases by 1\% or less, as shown in the column ``yoke only''.
In this column, $Q$ is determined by two losses, the losses of the RF-shield circuit and the magnetic loss in the yoke.
Since both act as resistances in series in the resonant loop, the relation of the quality factors can be written as $1/Q = 1/Q_\mathrm{circuit} + 1/Q_\mathrm{yoke}$. The value of $Q$ for lossless iron in this column gives $Q_\mathrm{circuit} = 61$, which happens to be close to the
free-space value of 63 in Sec.~\ref{sec:resfree}. With the values of $Q$ in the same column for $\mu_r = 1000$, $Q_\mathrm{yoke} = 1/(1/Q - 1/Q_\mathrm{circuit})$
is about 145 for $k_\mathrm{exc} = 1$ and 59 for $k_\mathrm{exc} = 2.5$. The yoke
accounts for about 30\% of the total loss for $k_\mathrm{exc} = 1$ and about 50\% for $k_\mathrm{exc} = 2.5$, and the rest is mainly the loss in the
capacitors. With the loss angle $\tan\delta = \mu''/\mu'$, the quality factor
due to the loss in the yoke alone is given by
\begin{equation}
\frac{1}{Q_\mathrm{yoke}} = \frac{W_\mathrm{iron}}{W}\tan\delta,
\label{eq:fill}
\end{equation}
where $W_\mathrm{iron}$ is the magnetic energy stored in the iron and $W$ is the total magnetic energy of the resonance. As shown in Sec.~\ref{sec:lam}, the loss
angle of the laminations at the resonance frequency is close to $45^\circ$ ($\tan\delta \approx 1$) for $k_\mathrm{exc} = 1$, and Eq.~(\ref{eq:fill}) with $Q_\mathrm{yoke} = 145$ gives $W_\mathrm{iron}/W \approx 0.7\%$.
For $k_\mathrm{exc} = 2.5$, both $\tan\delta$ and $1/Q_\mathrm{yoke}$ are about 2.5 times larger, and $W_\mathrm{iron}/W$ is the same. Since $|\mu_\mathrm{iron}| \approx 350$, the field strength $H = B/(\mu_0|\mu_\mathrm{iron}|)$ in the iron is small, and most of the
magnetic energy is stored inside the RF shield and in the 4.7~mm gap between the strips and the pole tips.

In the 2D calculation, the coils are assumed to be perfect conductors that exclude the flux, with no current through the coil terminals since the
coil reactance at 200~kHz is large (11~k$\Omega$ for EQ and 0.8~k$\Omega$ for RQ). Since the flux of the RF shield crosses the 4.7~mm gap into the pole tips,
returns through the iron and hardly enters the spaces between the poles, the coils change $f_r$ and $Q$ only by $+0.02\%$ and $-0.6\%$, respectively.

\section{Comparison with the resistive wall}\label{sec:beam}
The transverse resistive-wall impedance of the whole ring is evaluated with Métral's formula for a wall of finite thickness~\cite{Metral05}.
The wall over the whole circumference of 1567.5~m is assumed to be made of stainless steel, which is the main material of the MR vacuum chambers,
with a conductivity of $1.35\times10^6$~S/m, an inner radius of 62~mm and a thickness of 3~mm.
It is drawn as the black line in Fig.~\ref{fig:ReZ}.
The 206~kHz resonant peaks of $\mathrm{Re}\,Z_\perp$ per chamber plotted in Fig.~\ref{fig:ReZ} are comparable to the resistive-wall impedance of the whole ring at the
same frequency, and the three chambers together exceed it by a factor of 3--4.

\begin{table}[tb]
\caption{$\mathrm{Re}\,Z_\perp$ [k$\Omega$/m] at the unstable sidebands in slow extraction for one chamber ($\mu_r = 1000$, end plates) and for the resistive wall of the whole ring.}
\label{tab:sidebands}

\centering
\small\setlength{\tabcolsep}{4pt}
\begin{tabular}{lccrccc}
\toprule
 & & & & \multicolumn{3}{c}{$\mathrm{Re}\,Z_\perp$ [k$\Omega$/m]} \\
\cmidrule(lr){5-7}
 & $\nu$ & $n$ & $f$ [kHz] & $k_\mathrm{exc} = 1$ & $k_\mathrm{exc} = 2.5$ & resistive wall \\
\midrule
SX horizontal & 22.29 & 23 & 135.7 & 0.58 & 0.88 & 447 \\
              &       & 24 & 326.9 & 0.27 & 0.33 & 293 \\
SX vertical   & 20.80 & 21 & 38.2  & 0.16 & 0.19 & 777 \\
              &       & 22 & 229.4 & 5.53 & 7.67 & 350 \\
\bottomrule
\end{tabular}

\end{table}

We consider the transverse dipole instability of a coasting beam at 30~GeV, as during the debunching and the slow extraction.
Its growth rate is proportional to $\mathrm{Re}\,Z_\perp$ at the slow-wave sidebands $f = (n-\nu)f_\mathrm{rev}$ with $n > \nu$~\cite{Chao}, which
we refer to as the unstable sidebands. For the revolution frequency $f_\mathrm{rev} = 191.17$~kHz and the slow-extraction tunes
$(\nu_x,\nu_y) = (22.29, 20.80)$, the unstable sidebands are shown by the dotted lines in Fig.~\ref{fig:ReZ}. The resistive-wall impedance and the impedance of one
spill-feedback-quadrupole chamber at these sidebands are listed in Table~\ref{tab:sidebands}. The resonance peaks lie
between the sidebands, and the resistive wall dominates at all of them.

For bunched beams, the head-tail and coupled-bunch modes sample the impedance at the betatron sidebands, but each sideband is weighted by the
bunch spectrum shifted by the chromatic frequency~\cite{Chao}. The resonance impedance can therefore contribute depending on the chromaticity, the bunch
length and the head-tail mode. The evaluation of these instabilities is beyond the main purpose of this paper, and they are not discussed further.

\begin{table}[tb]
\caption{Resonance parameters of one chamber for different RF-shield materials and lamination thicknesses ($\mu_r = 1000$, SUS304 end plates).}
\label{tab:disc}
\centering
\small\setlength{\tabcolsep}{4pt}
\begin{tabular}{lcccc}
\toprule
 & $k_\mathrm{exc}$ & $f_r$ [kHz] & $Q$ & $R_\perp/Q$ [k$\Omega$/m] \\
\midrule
reference (copper RF shield, laminations 0.1~mm) & 1 & 205.8 & 42.9 & 12.21 \\
 & 2.5 & 205.9 & 29.9 & 12.17 \\
\midrule
stainless-steel RF shield & 1 & 203.7 & 12.7 & 12.43 \\
 & 2.5 & 203.7 & 11.2 & 12.43 \\
\midrule
laminations 0.35~mm & 1 & 207.6 & 26.7 & 11.98 \\
 & 2.5 & 208.4 & 14.9 & 11.92 \\
\bottomrule
\end{tabular}
\end{table}

\section{Discussion}\label{sec:discussion}
The quality factor of the resonance depends on the material of the RF shield and on the thickness of the laminations. The ISIS synchrotron uses
RF screens of stainless-steel wires and magnets with laminations 0.35~mm thick~\cite{HB23}, and we examine how these two choices change the resonance in our case.
In the final model, we replaced the copper RF shield by stainless steel and the 0.1~mm laminations by 0.35~mm ones, keeping the geometry, $\mu_r = 1000$
and the SUS304 end plates. The results are summarized in Table~\ref{tab:disc}.

\subsection{Material of the RF shield}
We replaced the copper of the strips and the rings by stainless steel ($\rho = 7.2\times10^{-7}~\Omega$m), whose resistivity is 42 times that of copper.
Since the skin depth increases, e.g., from 0.15~mm in copper to 0.94~mm in stainless steel at 200~kHz, the current flows through a larger part
of the cross section, and the increase of the effective resistance of the strips and rings is moderated to a factor of 20--30. Due to the increased loss in the RF shield,
$Q$ drops from 30--43 to 11--13 (Table~\ref{tab:disc}). The loss in the RF shield is now much larger than that in the iron. The resonance frequency and $R_\perp/Q$ change by only a few percent.

\subsection{Thickness of the laminations}
With laminations 0.35~mm thick instead of 0.1~mm (assuming the same magnetic properties and resistivity), $\mu_\mathrm{lam}$ at 200~kHz becomes $73-73j$ instead of $257-272j$ for
$k_\mathrm{exc} = 1$ as in Sec.~\ref{sec:lam}. Since $|\mu_\mathrm{lam}|$ is proportional to $\mu_r\delta/d$ in the skin-limited regime at the frequencies of interest while the loss angle stays close to $45^\circ$, the
energy stored in the iron, and hence the loss, increases approximately in proportion to $d$. For example, for $k_\mathrm{exc} = 1$, $Q_\mathrm{yoke}$ decreases from 145 to 47
and $Q$ from 43 to 27.

\subsection{Excitation of the magnets}
The calculations so far have assumed that the magnets are not excited. During slow extraction, however, EQ/RQ are excited, and the field of the RF shield is
superposed on a bias field in the iron. Since the magnet current changes at frequencies much lower than the resonance frequency, the bias field acts as a
static field, which may lower $\mu_r$ at the resonance frequency. For EQ, the pole-tip field is weak, about 70~mT (Sec.~\ref{sec:magnets}), and the flux density in the iron is of the
same order, which would be a few percent of saturation. The decrease of $\mu_r$ is therefore expected to be moderate and is covered by the case of $\mu_r = 500$, for which
$f_r$ and $R_\perp/Q$ hardly change and $Q$ decreases by less than 10\% (Sec.~\ref{sec:resyoke}). For main lattice magnets operated at higher fields,
the decrease of $\mu_r$ may need to be taken into account.

\section{Conclusion}

We evaluated the low-frequency transverse beam coupling impedance due to the RF shields with lumped capacitors on the ceramic chambers in the
J-PARC MR spill-feedback quadrupoles. The coupling impedance was analyzed with a magneto-quasi-static PEEC model of the RF shield, without the magnet and combined with a 2D field solution of
the laminated magnet yoke. The analysis shows a low-$Q$ resonant impedance at a few hundred kHz, in the same frequency range as that observed at ISIS. Without the magnet, the PEEC model gives a resonance at $f_r \approx 252$~kHz with $R_\perp/Q \approx 7.8$~k$\Omega$/m and $Q \approx 63$ per chamber. Since the domain-wall effects in the laminations (the excess loss and the decrease of $\mu_r$ at high frequency) at the frequency of interest are uncertain, they were evaluated with ranges of 1 to 2.5 for the ratio $k_\mathrm{exc}$ of the measured iron loss to the classical eddy-current loss and 500 to 1000 for the relative permeability $\mu_r$. 
Including the effect of the
laminated magnet, the PEEC analysis predicts a resonance at $f_r \approx 206$~kHz with $R_\perp/Q \approx 12$~k$\Omega$/m and $Q \approx 27$--43 per chamber.
The yoke lowers $f_r$ by 18\% and raises $R_\perp/Q$ by 57\% with respect to the case without the magnet. Therefore, the effect of the laminated
magnet cannot be neglected for EQ/RQ. The quality factor is determined mainly by the
loss in the capacitors and by the magnetic loss in the 0.1~mm laminations. The non-uniform current distribution across the strip width also increases the loss. The copper clamps and the titanium sleeves at the chamber ends raise $f_r$ by 5\%. The end faces of the yoke are shielded from the field of the RF shield by the stainless-steel end plates, and the coils have a negligible effect.

The value of $\mathrm{Re}\,Z_\perp$ per chamber at the resonant peak at about 206~kHz is comparable to the resistive-wall impedance of the whole
ring at the same frequency, and the three chambers together exceed it by a factor of 3--4. For a coasting beam,
as during the debunching and the slow extraction, the resonance peaks lie between the unstable sidebands, where the resistive wall dominates.
For bunched beams, the head-tail and coupled-bunch instabilities depend on the chromaticity and the bunch length. In the actual fast-extraction operation, transverse instabilities are suppressed by the transverse feedback system.

The method developed here can be applied to the RF shields in other magnets, including dipole magnets, of existing machines such as ISIS, CSNS and the J-PARC Rapid Cycling Synchrotron~\cite{Kinsho05}, and to the design of future rapid-cycling
synchrotrons. The Julia code of the PEEC model takes about 1.3~s per frequency on a laptop computer and minutes to tens of minutes for one resonance curve.

\section*{Acknowledgments}
This work was motivated by the ISIS--KEK collaboration meeting held on 28--29 September 2026, and the author thanks its organizers and the participants
from ISIS and KEK. The fabrication of the ceramic chambers and their RF shields was conducted by the J-PARC MR vacuum group. The author thanks the J-PARC MR
slow-extraction and beam-instability groups for their support.

\section*{Data availability}
The Julia scripts used in this work will be made available on request.

\end{document}